\documentclass[prb,aps,english,twocolumn,notitlepage]{revtex4-1}
\usepackage[T1]{fontenc}  \usepackage[latin1]{inputenc}
 \usepackage{babel} 
 \usepackage{epsfig,epsf,psfrag} 
\usepackage{graphicx} 
\usepackage{subfigure}
\usepackage{pslatex}
\usepackage{epic,eepic} \usepackage{bbold}
\usepackage{color,pstcol}\usepackage{amsmath}
\usepackage{pstricks} 
\usepackage{fancyhdr}  \parindent0em  

 \vspace{-10.0cm}  
\begin{document} \title{Disordered contacts can localize chiral edge electrons} 
  \author{Arjun Mani}
 \author{Colin Benjamin} \email{colin.nano@gmail.com}\affiliation{School of Physical Sciences, National Institute of Science Education \& Research, HBNI, Jatni-752050, India}
\begin{abstract}
Chiral integer quantum Hall (QH) edge modes are immune to backscattering and therefore are non-localized and show a vanishing longitudinal as well as non-local resistance along with quantized 2-terminal and Hall resistance even in presence of sample disorder. However, this is not the case for contact disorder, which refers to the possibility that a contact can reflect edge modes either partially or fully. This paper shows that when all contacts are disordered in a N-terminal quantum  Hall bar, then transport via chiral QH edge modes can have a significant localization correction. The Hall and 2-terminal resistance in an N-terminal quantum Hall sample deviate from their values derived while neglecting the phase acquired at disordered contacts, and this deviation is called the quantum localization correction. This correction term increases with the increase of disorderedness of contacts but decreases with the increase in number of contacts in an N-terminal Hall bar. The presence of inelastic scattering, however, can completely destroy the quantum localization correction.  
\end{abstract}
 \maketitle
\section{Introduction}
One of the most important discoveries of the twentieth century is the quantum Hall effect in 2DEG owing to its robust dissipation less 1D edge modes \cite{goerbig, asboth}. These 1D edge modes, robust to sample disorder, are observed in quantum Hall (QH) systems at low temperatures in presence of a perpendicular magnetic field\cite{buti, datta}. QH edge modes are chiral, i.e., at one edge of the system electrons are moving in one direction and at the other edge electrons are moving in opposite direction. Due to the topological character of these edge modes, the Hall conductance in QH case is quantized to $\frac{2e^2}{h}$. Determining the Hall, longitudinal, 2-terminal and nonlocal conductances/resistances in QH samples is done by taking recourse to Landauer-Buttiker (L-B) formalism. According to this formalism, for a multi-terminal device at zero temperature, the current at contact $i$ is given as follows\cite{datta, buti, nikolajsen}:
\begin{eqnarray}
I_i&=&\sum_{\substack{j=1 \\ j\neq i}}^N[G_{ij}V_i-G_{ji}V_j]=\frac{2e^2}{h}\sum_{\substack{j=1 \\ j\neq i}}^N[T_{ji}V_i-T_{ij}V_j],\\
&& \text{with }T_{ij}=Tr[s_{ij}^\dagger s_{ij}]\nonumber
\end{eqnarray}
 \begin{figure}
   \centering {\includegraphics[width=0.4\textwidth]{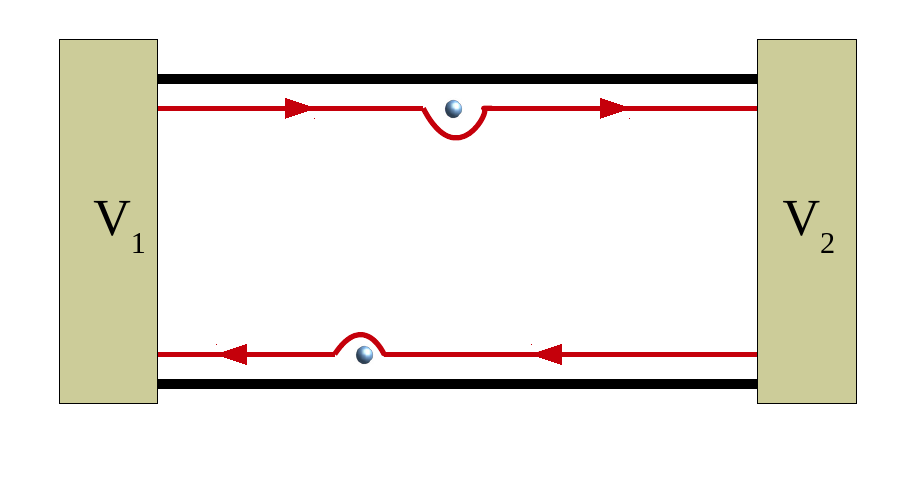}}\\
 \vspace{-.3cm}
\caption{Chiral edge modes immune to sample disorder.}
\end{figure}
where $T_{ij}$ is the transmission probability for an electron from contact $j$ to contact $i$, $V_i$ is the potential bias at contact $i$, and $s_{ij}$ are the elements of the scattering matrix $S$ of the $N$ terminal QH system. These quantum Hall (QH) edge modes being chiral and dissipation-less are highly promising candidates for use in low power information processing \cite{asboth, buti}. It is well known that transport via these QH edge modes is robust against disorder. Thus, the Hall and local or 2-terminal resistance calculated for a QH sample are invariant to disorder. In this work, however, we predict that a crucial distinction arises between a quantum Hall sample with all $N$ contacts disordered and that with less than $N$ contacts disordered (say $1, 2$ or even $N-1$ contacts disordered). We predict that for a sample with all N contacts disordered a quantum localization correction term comes into being due to interference however with less than N contacts disordered there is no localization correction term. We identify this correction term to the resistances as quantum localization to distinguish it from weak localization which arises in quantum diffusive transport. The reason this was not predicted earlier was because there is a difference between the resistances derived via probabilities and that via scattering amplitudes as we explain in the next section. 

\begin{figure}[h]
 \centering \subfigure[]{\includegraphics[width=0.45\textwidth]{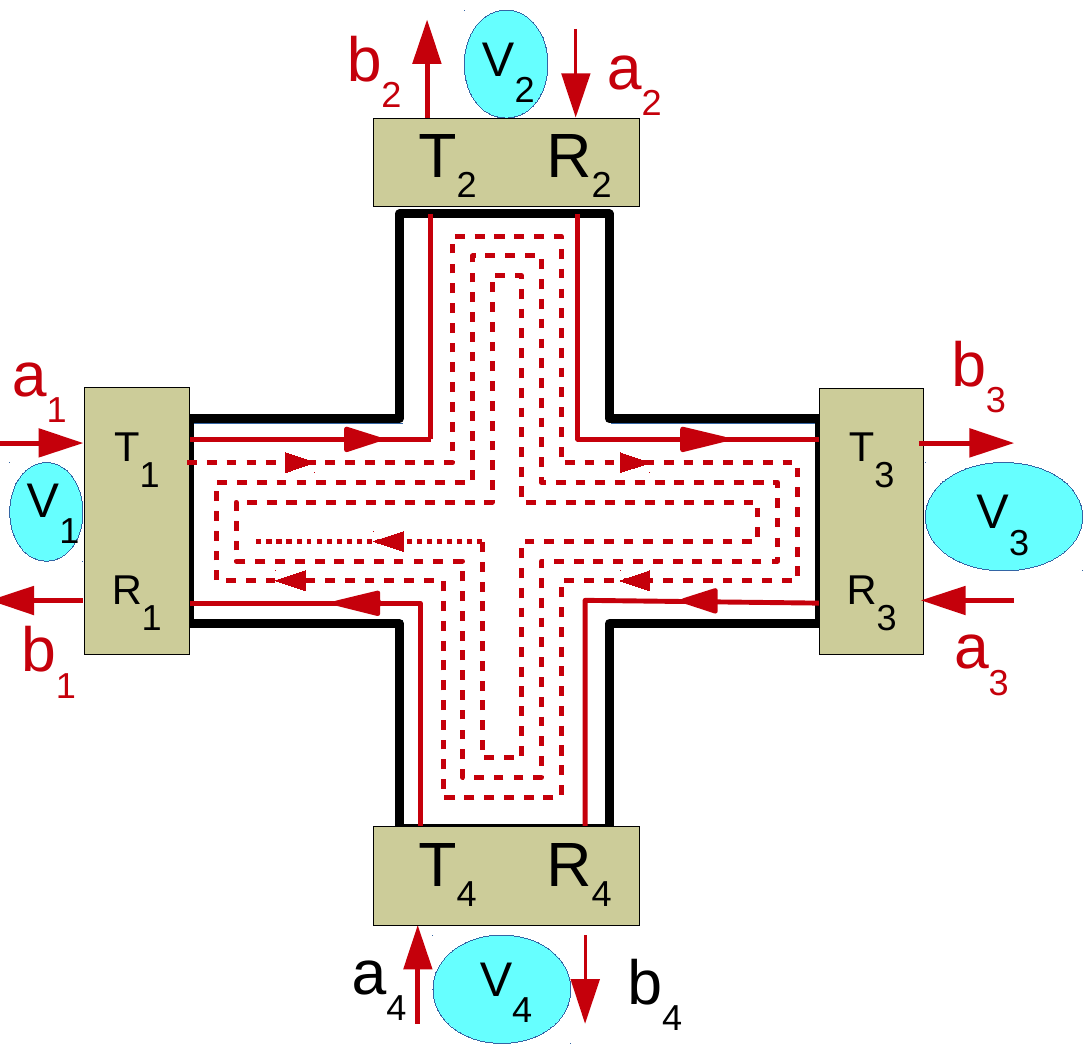}}
  \centering \subfigure[]{\includegraphics[width=0.45\textwidth]{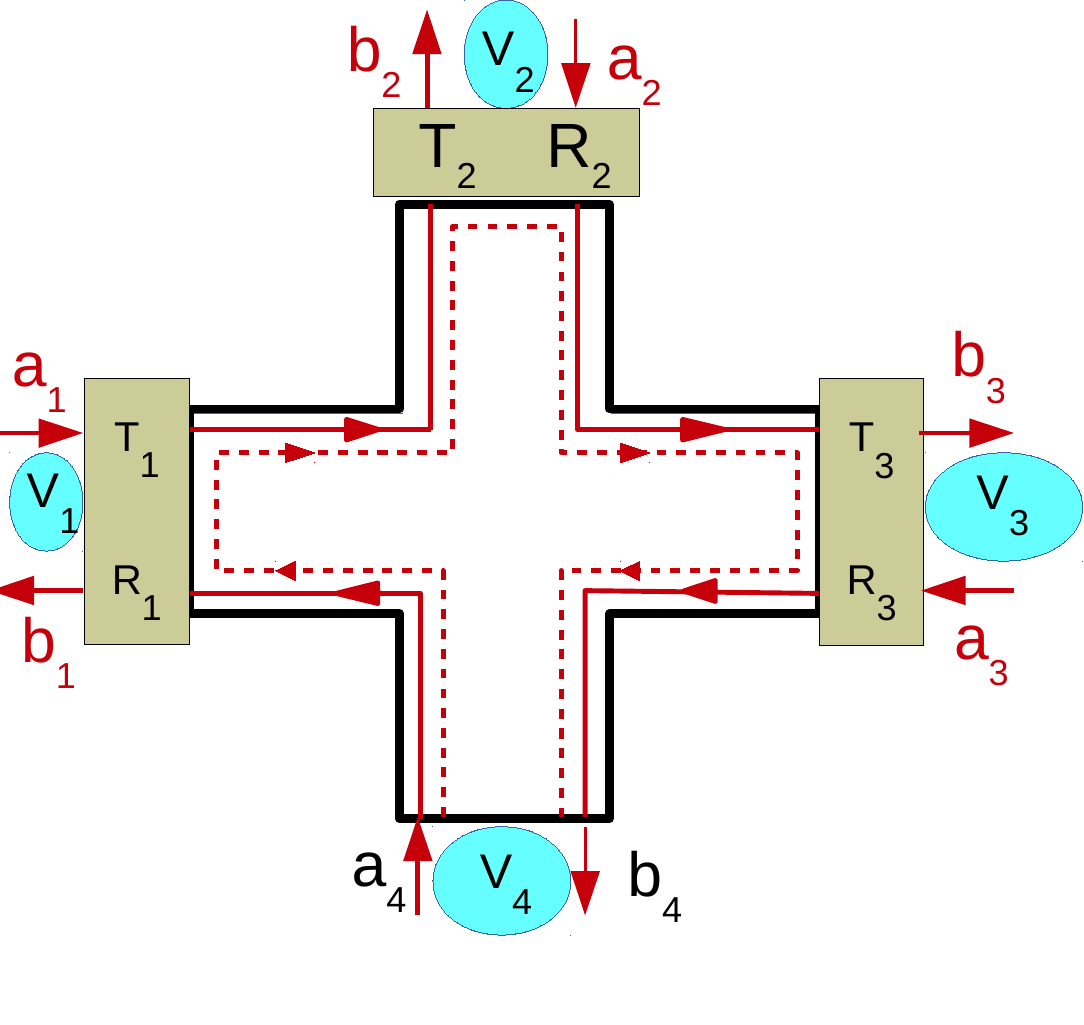}}\\
 \vspace{-.3cm}
\caption{4 terminal QH bar with (a) quantum localization correction when all contacts disordered, (b) no quantum localization correction for three contacts disordered.}
\end{figure}

 \begin{figure*}
 \centering \subfigure[]{\includegraphics[width=0.25\textwidth]{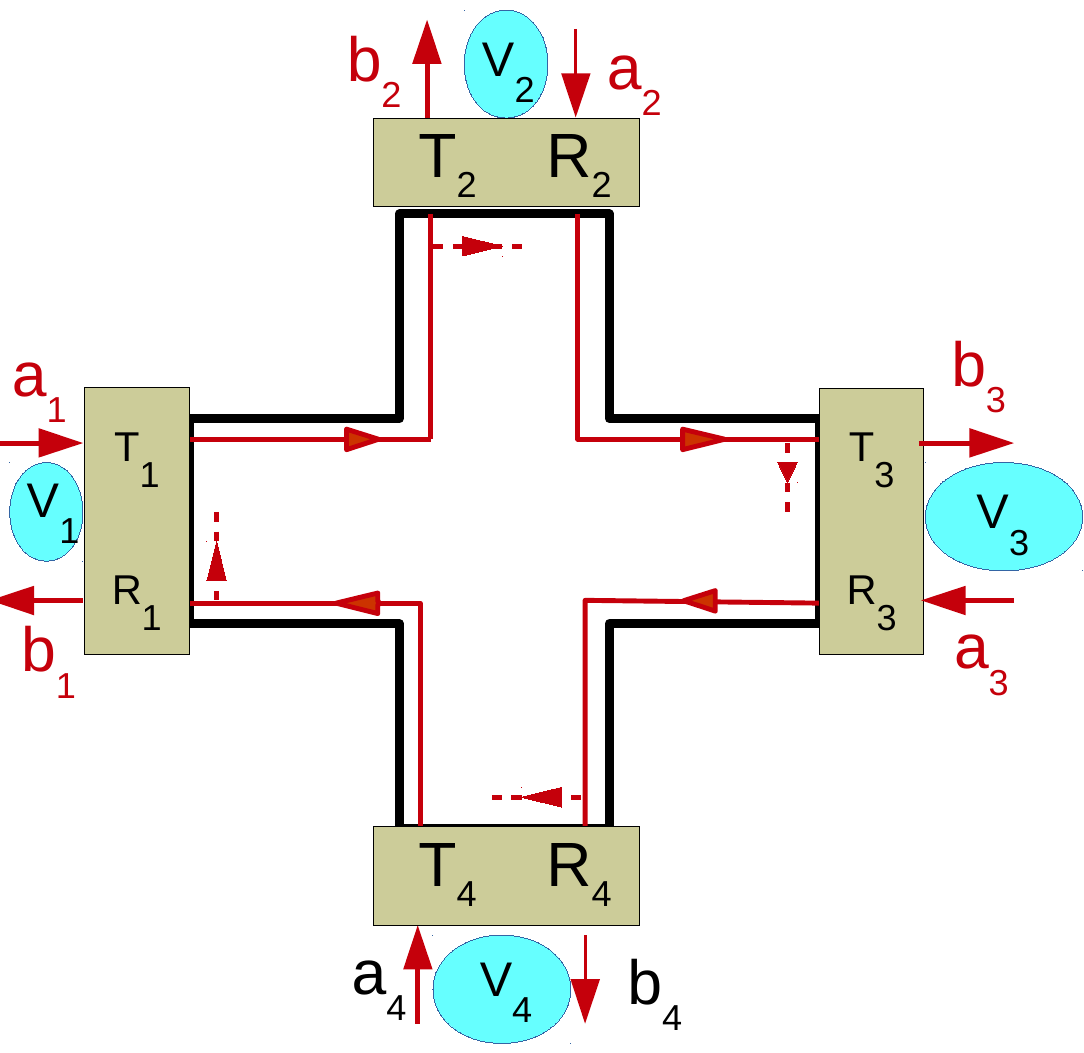}}
  \centering \subfigure[]{\includegraphics[width=0.33\textwidth]{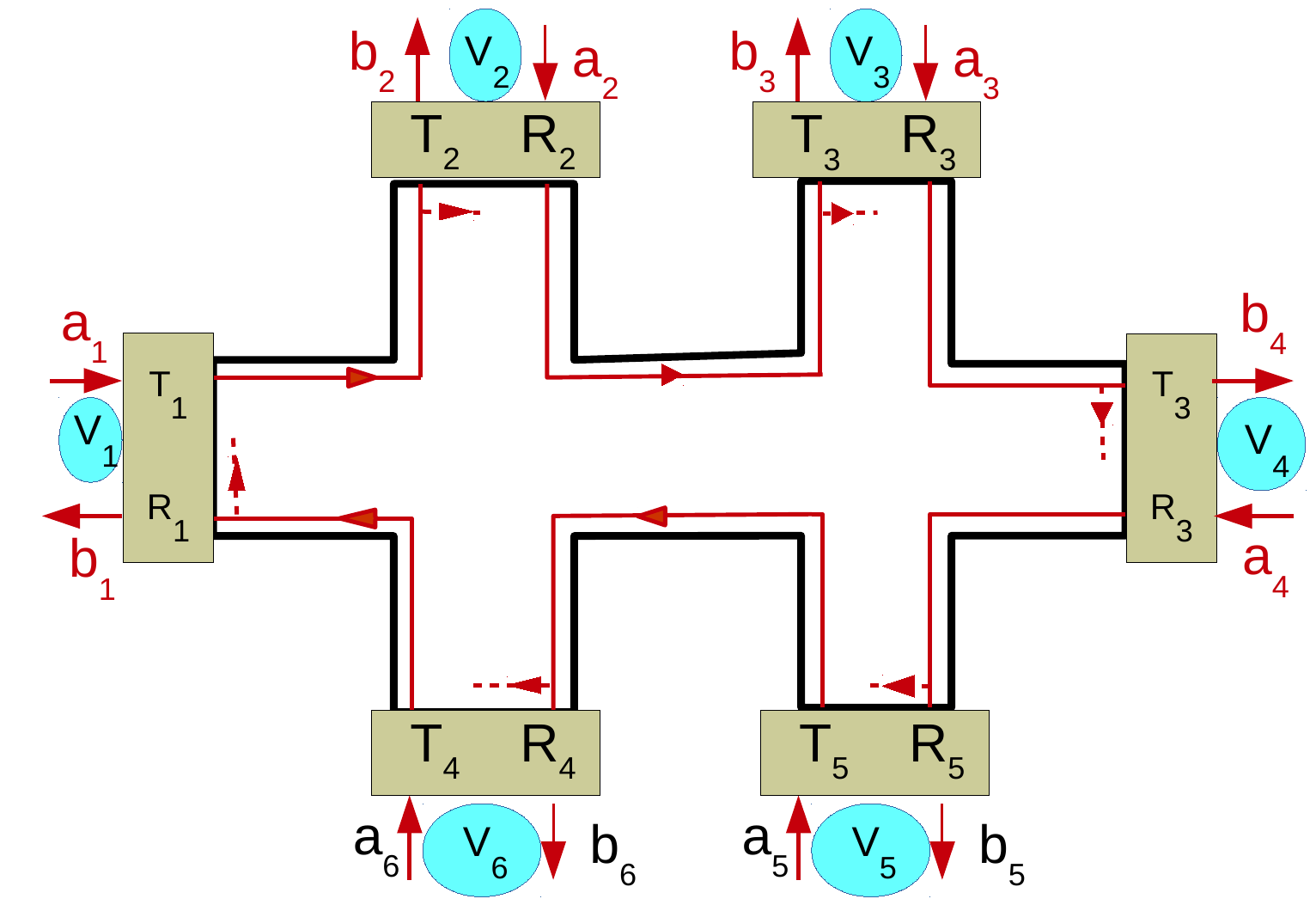}}
  \centering\subfigure[]{\includegraphics[width=.38\textwidth]{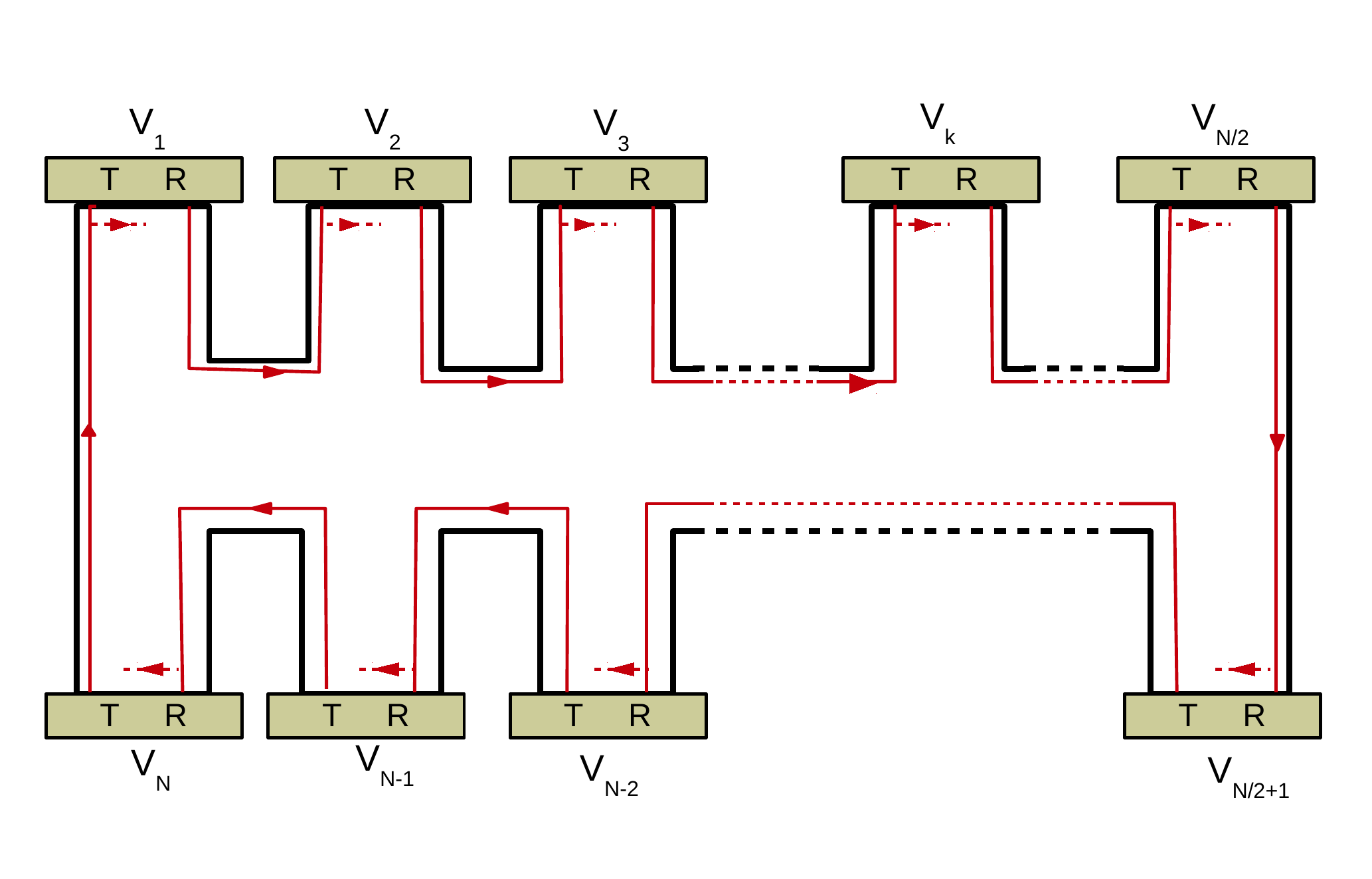}}
 \vspace{-.3cm}
\caption{(a) 4 terminal, (b) 6 terminal and (c) N terminal QH bar with all contacts disordered.}
\end{figure*}
\section{Motivation}
In the quantum quasi-ballistic or diffusive transport regime, it has been observed that the transmission probability of an electron through N number of scatterers depends on the number of scatterers, and thus, on the length ($l$) of the mesoscopic system. Due to this, localization of electrons, the resistance of the system increases exponentially with the system length ($l$), see Ref.~\cite{datta}. This is called strong or Anderson localization\cite{jian,jain}. When sample length $l\leq \xi$ (localization length) then the resistance of the system displays an universal behavior in that the resistance increases from the Ohmic result (increasing linearly with length) by the universal factor $\frac{h}{2e^2}$. This increase by the universal constant $\frac{h}{2e^2}$ is termed the weak localization correction to quantum transport. This weak localization effect has not been observed for chiral edge mode transport due to the robustness of edge modes to disorder \cite{buti}. If there is disorder in the sample, edge modes will bypass the disorder without affecting their transmission probabilities due to them being topologically protected, see Fig.~1. However, in this work we show if disorder is at a contact which can reflect the edge modes partially then a quantum localization correction can also appear for edge mode transport, when all contacts are disordered. We show that presence of disorder at all the contacts introduces backscattering within the system and generates multiple paths from one contact to another. This occurrence of multiple paths can be explained in more detail by alluding to Fig.~2(a). In Fig.~2(a), we see that an electron in the $a_1$ edge mode  can either transmit to the sample with  probability $T_1$ or it can reflect back to contact $1$ with probability $R_1$. Then, after entering into the sample, it can reach contact $3$ only after reflecting from contact $2$ with  probability $R_2$ and then transmitting into contact $3$ with probability $T_3$. Thus, the net transmission probability for an electron from contact $1$ to $3$ is $T_1R_2T_3$. However, this is just one path among the infinite number of paths available from contact $1$ to $3$. The electron from $a_1$ edge mode  can also reach contact $3$ by taking an alternative second path, such as: after transmitting into the sample with  probability $T_1$ it can reflect from contacts $2$, $3$, $4$ then reflect again at contacts $1$ and $2$ and then finally transmit into contact $3$ with probability $T_1R_1R_2R_3R_4R_1R_2T_3$. Further, the electron from $a_1$ edge mode can also reach contact $3$ by following a third path with multiple reflections at all the contacts and then a fourth path is possible which has many more reflections than the third path and so on, an infinite number of paths are possible from contact $1$ to $3$. Summing over all these infinite number of paths we get the total transmission probability from contact $1$ to $3$ as $T_{31}=\frac{T_1R_2T_3}{(1-R_1R_2R_3R_4)}$. We  see from the expression of $T_{31}$ that even when a single contact is not disordered, say contact $4$ (i.e., $R_4=0$) then $T_{31}=T_1R_2T_3$ which is equal to contribution from the first path. That means the contribution of all other paths to $T_{31}$ vanishes. This eventually makes the transmission probabilities and resistances dependent on the disorderedness of contacts. In this explanation of occurrence of multiple paths we neglected the phase acquired when an edge mode scatters from the contact. However, this phase is the reason for the localization correction. As because of the multiple paths available to an edge mode when all contacts are disordered, the phase acquired by the edge mode at each contact does not vanish when all the paths are summed, leading to a difference in the Resistances calculated by summing the amplitudes (without neglecting the phase) and the Resistances calculated by summing the probabilities while neglecting the phase. This has explained in more detail in section III.

The quantum localization correction to QH edge mode transport has never been studied before and is the main motivation of this work. Although, here we note that the quantum localization effect seen for edge modes is different from the universal weak localization correction observed in quantum diffusive transport. In quantum diffusive transport regime, the weak localization correction term is universal and equal to $\frac{h}{2e^2}$, while in our case it depends on the amount of disorder at the contacts. Again in our work we see quantum localization correction is inversely proportional to the number of contacts and is only present when all contacts are disordered, see Fig.~2(a). As when all contacts are disordered only then there would be an infinite number of paths from one contact to another (Fig.~2(a)), which eventually leads to a difference between the average resistances derived from scattering amplitudes $<R^{Amp}_X>$ (with $X$ being either the Hall or the 2-terminal) and resistances derived from probabilities $R^{}_X$, i.e., $<R^{Amp}_X>\neq R^{}_X$. Of course, $R_X^{}$ is derived neglecting the phase acquired by the electrons via scattering at the disordered contacts. If at least one contact is not disordered, see Fig.~2(b) (here contact $4$ is not disordered), then backscattering is absent, which leads to $<R^{Amp}_X>= R^{}_X$ thus the quantum localization correction vanishes. We elaborate on this in section III.

The manuscript is organized as follows: in section III, we deal with four and six terminal QH system with all contacts disordered and derive an expression for the quantum localization correction for N terminal system. In section IV, we study the effect of inelastic scattering on the quantum localization correction. Finally, we conclude with a table summarizing the main results of our paper.

 \section{Edge modes in Quantum Hall sample with disordered contacts} 
 \subsection{Four terminal QH bar with all contacts disordered}
A four terminal QH system is shown in Fig.~3(a) with all contacts disordered. Contacts $1$, $3$ are current probes and contacts $2$, $4$ are voltage probes, such that current through voltage probes $I_2=I_4=0$. A contact $i$ is disordered with disorder strength $D_i$ implies that the transmission probability of an electron through the contact gets reduced from unity to $(1-D_i)=T_i$, with $T_i+R_i=1$ ($R_i=D_i$ being the reflection probability at contact $i$). First we write down the scattering matrix $s_j$ at each contact $j$ separately relating the incoming wave ($a_j, a'_j$) to the outgoing wave ($b_j,b'_j$) at that particular contact $i$ and then derive the total scattering matrix $S$ of the system out of these contact scattering matrices $s_j$. The scattering matrix $s_j$ is defined as follows- 
 \begin{equation}
 s_j=\begin{pmatrix}
 r_je^{i\phi_j^r}&t_je^{i\phi_j^t}\\
 t_je^{i\phi_j^t}&r_je^{i\phi_j^r}
 \end{pmatrix},
 \end{equation}
where $r_j$ and $t_j$ are the reflection and transmission amplitudes respectively at contact $j$, $\phi^r_j$ and $\phi^t_j$ are the phases acquired by the electron by reflecting or transmitting at the disordered contact $j$. Since scattering matrix $s_j$ has to be unitary, i.e. $s_j^\dagger s_j=s_js_j^\dagger=I$ (where $I$ is Identity matrix). So in Eq.~(2)- $\phi_j^r=\phi_j^t-\pi/2=\phi_j$. Thus, the scattering matrix $s_j$ for each contact from Eq.~(2) is-
  \begin{equation}
 s_j=\begin{pmatrix}
 r_je^{i\phi_j}&it_je^{i\phi_j}\\
 it_je^{i\phi_j}&r_je^{i\phi_j}
 \end{pmatrix},
 \end{equation}
which connects the incoming wave to the outgoing wave via $(b_j,b_j')^T=s_j(a_j,a_j')^T$, with $j=1-4$. Each element of the total scattering matrix ($S$) can be calculated from these $s_j$ matrices in the following way: suppose an electron incoming in edge state ($a_1$) can reflect as $b_1$ edge state with amplitude $r_1e^{i\phi_1}$, but then, it can also follow a different path by transmitting through contact $1$ and then get reflected at contact $2$, then again get reflected at $3$ and then at $4$ after which it transmits through contact $1$ into $b_1$ state. The amplitude for this path is $it_1e^{i\phi_1}\times r_2e^{i\phi_2}\times r_3e^{i\phi_3}\times r_4e^{i\phi_4}\times it_1e^{i\phi_1} =-t_1^2r_2r_3r_4e^{i(2\phi_1+\phi_2+\phi_3+\phi_4)}$. Following this one can also have a third path with amplitude $-t_1^2r_1(r_2r_3r_4)^2e^{i(3\phi_1+2\phi_2+2\phi_3+2\phi_4)}$ and so on. Summing all these terms we get the scattering amplitude from contact $1$ to itself, as $(r_1-r_2r_3r_4e^{i\phi})e^{i\phi_1}/(1-r_1r_2r_3r_4e^{i\phi})$, with $\phi=\phi_1+\phi_2+\phi_3+\phi_4$. Similarly, rest of the elements of the total scattering matrix ($S$) of the four terminal QH system can be derived. The scattering matrix for the system in Fig.~3(a) is-
 {\footnotesize
\begin{equation} 
 S=\frac{1}{a}\left(\begin{smallmatrix}
( r_1-r_2r_3r_4e^{i\phi})e^{i\phi_1}&-t_1t_2r_3r_4e^{i\phi}&-t_1t_3r_4e^{i(\phi-\phi_2)}&-t_1t_4e^{i(\phi_1+\phi_4)}\\
 -t_1t_2e^{i(\phi_1+\phi_2)}&(r_2-r_1r_3r_4 e^{i\phi})e^{i\phi_2}&-t_2t_3r_1r_4e^{i\phi}&-t_2t_4r_1e^{i(\phi-\phi_3)}\\
 -t_1t_3r_2e^{i(\phi-\phi_4)}&-t_2t_3e^{i(\phi_2+\phi_3)}&(r_3-r_1r_2r_4e^{i\phi})e^{i\phi_3}&-t_3t_4r_1r_2e^{i\phi}\\
 -t_1t_4r_3r_2e^{i\phi}&-t_2t_4r_3e^{i(\phi-\phi_1)}&-t_3t_4e^{i(\phi_3+\phi_4)}&(r_4-r_1r_2r_3e^{i\phi})e^{i\phi_4}\end{smallmatrix}\right),
 \end{equation}
}where $a=1-r_1r_2r_3r_4e^{i\phi}$. This scattering matrix $S$ relates the incoming edge states to the outgoing edge states (see Fig.~3(a)) of the system by the relation $(b_1,b_2,b_3,b_4)^T=S (a_1,a_2,a_3,a_4)^T$. Unitarity of the scattering matrix $S$ shows the conservation of current within the system. To calculate the current at each of these contacts, we need to derive the transmission probabilities $T^{}_{ij}$ between these contacts by following Eq.~(4).\\
The conductance matrix $G$ of the system derived from the scattering matrix $S$ of Eq.~(4), following Eq.~(1), is-
 {\scriptsize
\begin{equation} 
 G=\frac{2e^2}{h}\frac{1}{a'}\begin{pmatrix}
(1-R_2R_3R_4)T_1&-T_1T_2R_3R_4&-T_1T_3T_4&-T_1T_4\\
 -T_1T_2&(1-R_1R_3R_4)T_2&-T_2T_3R_1R_4&-T_2T_4R_1\\
 -T_1T_3R_2&-T_2T_3&(1-R_1R_2R_4)T_3&-T_3T_4R_1R_2\\
 -T_1T_4R_3R_2&-T_2T_4R_3&-T_3T_4&(1-R_1R_2R_3)T_4\end{pmatrix},
 \end{equation}
}where $a'=(1+R_1R_2R_3R_4-2\sqrt{R_1R_2R_3R_4}\cos\phi)$, $T_i=1-D_i=|t_i|^2$ and $R_i=D_i=|r_i|^2$ for $i=1-4$ with $r_i$ and $t_i$ are defined as shown in Eqs.~(2,3). The conductance matrix $G$ connects the currents and voltages at each of the contacts by the relation $(I_1,I_2,I_3,I_4)^T=G(V_1,V_2,V_3,V_4)^T$. Since the current through voltage probe $2$ and $4$ are zero, so $I_2=I_4=0$, and choosing the reference potential $V_3=0$ we get the potentials $V_2$ and $V_4$ in terms of $V_1$. The Hall resistance $R^{Amp}_H=R_{13,24}=\frac{(V_2-V_4)}{I_1}$, 2-terminal resistance $R^{Amp}_{2T}=R_{13,13}=\frac{(V_1-V_3)}{I_1}$, and non-local resistance $R^{Amp}_{NL}=R_{12,43}=\frac{(V_4-V_3)}{I_1}$ (to calculate the non-local resistance only we have to consider contacts $1,2$ are current probes and contacts $3,4$ are voltage probes) become-
\begin{eqnarray}
R^{Amp}_H&=&\frac{h}{2e^2}\frac{1+D_1D_2D_3D_4-2\sqrt{D_1D_2D_3D_4}\cos\phi}{1-D_1D_2D_3D_4},\nonumber\\
R^{Amp}_{2T}&=&\frac{h}{2e^2}\frac{(1-D_1D_3)(1+D_1D_2D_3D_4-2\sqrt{D_1D_2D_3D_4}\cos\phi)}{(1-D_1)(1-D_3)(1-D_1D_2D_3D_4)},\nonumber\\
R^{Amp}_{NL}&=&0.
\end{eqnarray}
After averaging over the phase shift $\phi$ acquired by the electron via multiple scatterings at all disordered contacts we get the mean Hall, 2-terminal and nonlocal resistances as follows- 
\begin{eqnarray}
\langle R^{Amp}_X\rangle&=&\frac{1}{2\pi}\int_{0}^{2\pi}R^{Amp}_X d\phi, \quad\text{with }X=H,2T,NL,\nonumber\\
\text{given as-}&&\nonumber\\
\langle R^{Amp}_H\rangle&=&\frac{h}{2e^2}\frac{1+D_1D_2D_3D_4}{1-D_1D_2D_3D_4},\nonumber\\
\langle R^{Amp}_{2T}\rangle&=&\frac{h}{2e^2}\frac{(1-D_1D_3)(1+D_1D_2D_3D_4)}{(1-D_1)(1-D_3)(1-D_1D_2D_3D_4)},\nonumber\\
\langle R^{Amp}_{NL}\rangle&=&0.
\end{eqnarray}

We see that the Hall and 2-terminal resistance lose their quantization, although, the non-local resistance remains unaffected by disorder due to the chiral nature of QH edge mode transport. \\
To calculate the quantum localization correction, we calculate the Hall, 2-terminal and non-local resistances via probabilities summing the paths and neglecting the amplitude and phase. The conductance matrix $G$ defined by probabilities of reflection/transmission at each contact is given as\cite{arjun, arjun1}- 

{\scriptsize
\begin{equation} 
 G=\frac{2e^2}{h}\frac{1}{a''}\begin{pmatrix}
(1-R_2R_3R_4)T_1&-T_1T_2R_3R_4&-T_1T_3T_4&-T_1T_4\\
 -T_1T_2&(1-R_1R_3R_4)T_2&-T_2T_3R_1R_4&-T_2T_4R_1\\
 -T_1T_3R_2&-T_2T_3&(1-R_1R_2R_4)T_3&-T_3T_4R_1R_2\\
 -T_1T_4R_3R_2&-T_2T_4R_3&-T_3T_4&(1-R_1R_2R_3)T_4\end{pmatrix},
 \end{equation}} where $a''=(1-R_1R_2R_3R_4)$. Again, setting the current through voltage probes $2,4$ equal to zero, and choosing the reference potential $V_3=0$ we get the potentials $V_2$ and $V_4$ in terms of $V_1$. We get the Hall resistance $R_H$, 2-terminal resistance $R^{}_{2T}$, and nonlocal resistance $R^{}_{NL}$ via probabilities as-
\begin{eqnarray}
R^{}_H&=&\frac{h}{2e^2},
\text{ } R^{}_{2T}=\frac{h}{2e^2}\frac{(1-D_1D_3)}{(1-D_1)(1-D_3)},\text{ } R^{}_{NL}=0.
\end{eqnarray}
Here, we get the quantization of the Hall resistance back. The difference between resistance calculated via amplitudes and probabilities gives quantum localization correction ($R^{QL}_X=\langle R^{Amp}_X\rangle-R^{}_X$, with $X=H, 2T, NL$) for Hall, 2-terminal and non-local resistance. The quantum localization correction thus is-
\begin{eqnarray}
R^{QL}_H&=&\frac{h}{2e^2}\frac{2D_1D_2D_3D_4}{1-D_1D_2D_3D_4},\nonumber\\
R^{QL}_{2T}&=&\frac{h}{2e^2}\frac{2D_1D_2D_3D_4(1-D_1D_3)}{(1-D_1)(1-D_3)(1-D_1D_2D_3D_4)},\nonumber\\
R^{QL}_{NL}&=&0.
\end{eqnarray}
Note that if not all contacts are disordered then the quantum localization correction vanishes. This can be seen from Eqs.~(7, 9) and Eq.~(10), i.e., if we consider some contact $i$ is not disordered, i.e., $D_i=0$ (for $i=1/2/3/4$) then $R_X^{QL}=0$ and thus $\langle R_X^{Amp}\rangle=R_X^{}$, with $X=H, 2T, NL$. This condition holds for any number of terminals as we show below. Thus the resistance calculated via probabilities and via amplitudes are identical.

 \subsection{Six terminal system with all contacts disordered}
A six terminal QH system is shown in Fig.~3(b) with all contacts disordered. Contacts $1$ and $4$ are current probes and contacts $2, 3, 5, 6$ are voltage probes, so, current through these contacts are set to zero, i.e., $I_2=I_3=I_5=I_6=0$. The scattering matrix for the system in Fig.~3(b) is-
{\footnotesize
\begin{equation} 
 S=\frac{1}{b}\left(\begin{smallmatrix}
 (r-r^5e^{i\phi})e^{i\phi_1}&-t^2r^4e^{i\phi}&-t^2r^3e^{i\phi_{34561}}&-t^2r^2e^{i\phi_{4561}}&-t^2re^{i\phi_{561}}&-t^2e^{i\phi_{61}}\\
 -t^2e^{i\phi_{12}}&(r-r^5e^{i\phi})e^{i\phi_2}&-t^2r^4e^{i\phi}&-t^2r^3e^{i\phi_{45612}}&-t^2r^2e^{i\phi_{5612}}&-t^2re^{i\phi_{612}}\\
 -t^2re^{i\phi_{123}}&-t^2e^{i\phi_{23}}&(r-r^5e^{i\phi})e^{i\phi_3}&-t^2r^4e^{i\phi}&-t^2r^3e^{i\phi_{56123}}&-t^2r^2e^{i\phi_{6123}}\\
 -t^2r^2e^{i\phi_{1234}}&-t^2re^{i\phi_{234}}&-t^2e^{i\phi_{34}}&(r-r^5e^{i\phi})e^{i\phi_4}&-t^2r^4e^{i\phi}&-t^2r^3e^{i\phi_{61234}}\\
 -t^2r^3e^{i\phi_{12345}}&-t^2r^2e^{i\phi_{2345}}&-t^2re^{i\phi_{345}}&-t^2e^{i\phi_{45}}&(r-r^5e^{i\phi})e^{i\phi_5}&-t^2r^4e^{i\phi}\\
 -t^2r^4e^{i\phi}&-t^2r^3e^{i\phi_{23456}}&-t^2r^2e^{i\phi_{3456}}&-t^2re^{i\phi_{456}}&-t^2e^{i\phi_{56}}&(r-r^5e^{i\phi})e^{i\phi_6}\end{smallmatrix}\right),
 \end{equation}
}where $b=1-r^6e^{i\phi}$. Here, $\phi=\phi_1+\phi_2+\phi_3+\phi_4+\phi_5+\phi_6$, the sum of all scattering phases acquired at each disordered contact and $\phi_{ij..m}=\phi_i+\phi_j+...+\phi_m$. This scattering matrix $S$ relates incoming waves to outgoing waves (see Fig.~3(b)) of the system by the relation:\\ $(b_1,b_2,b_3,b_4,b_5,b_6)^T=S(a_1,a_2,a_3,a_4,a_5,a_6)^T$. Unitarity of the scattering matrix $S$ implies the conservation of current in the system. The conductance matrix $G$ of the system derived from the scattering matrix $S$ in Eq.~(11) following from Eq.~(1), is-
{
\begin{equation} 
 G=\frac{2e^2}{h}\frac{1}{b'}\left(\begin{smallmatrix}
(1-R^5)T&-T^2R^4&-T^2R^3&-T^2R^2&-T^2R&-T^2\\
 -T^2&(1-R^5)T&-T^2R^4&-T^2R^3&-T^2R^2&-T^2R\\
 -T^2R&-T^2&(1-R^5)T&-T^2R^4&-T^2R^3&-T^2R^2\\
 -T^2R^2&-T^2R&-T^2&(1-R^5)T&-T^2R^4&-T^2R^3\\
 -T^2R^3&-T^2R^2&-T^2R&-T^2&(1-R^5)T&-T^2R^4\\
 -T^2R^4&-T^2R^3&-T^2R^2&-T^2R&-T^2&(1-R^5)T\end{smallmatrix}\right),
 \end{equation}
}where $b'=(1+R^6-2R^3\cos\phi)$. Since the current through voltage probes $2,3,5$ and $6$ are zero, so $I_2=I_3=I_5=I_6=0$, and choosing the reference potential $V_4=0$ we get the potentials $V_2$, $V_3$, $V_5$ and $V_6$ in terms of $V_1$. So, the Hall resistance $R^{Amp}_H=R_{14,26}=\frac{(V_2-V_6)}{I_1}$, 2-terminal resistance $R^{Amp}_{2T}=R_{14,14}=\frac{(V_1-V_4)}{I_1}$, longitudinal resistance $R^{Amp}_L=R_{14,23}=\frac{(V_2-V_3)}{I_1}$ and non-local resistance $R^{Amp}_{NL}=R_{12,54}=\frac{(V_5-V_4)}{I_1}$ (to calculate the non-local resistance only we have to consider contacts $1,2$ are current probes and contacts $3,4,5,6$ are voltage probes) becomes-
\begin{eqnarray}
R^{Amp}_H&=&\frac{h}{2e^2}\frac{1+D^6-2D^3\cos\phi}{1-D^6},\nonumber\\
R^{Amp}_{2T}&=&\frac{h}{2e^2}\frac{(1+D)(1+D^6-2D^3\cos\phi)}{(1-D)(1-D^6)},\nonumber\\
R^{Amp}_L&=&R^{Amp}_{NL}=0.
\end{eqnarray}
Here, for simplicity we have considered that all the contacts are equally disordered, i.e., $D_i=D$ (for $i=1-6$). After averaging over the phase shift $\phi$ acquired by the electron in a round trip from $0$ to $2\pi$ we get, 
\begin{eqnarray}
\langle R^{Amp}_H\rangle&=&\frac{h}{2e^2}\frac{1+D^6}{1-D^6},\quad
\langle R^{Amp}_{2T}\rangle=\frac{h}{2e^2}\frac{(1+D)(1+D^6)}{(1-D)(1-D^6)},\nonumber\\
\langle R^{Amp}_L\rangle&=&\langle R^{Amp}_{NL}\rangle=0.
\end{eqnarray}
The longitudinal and non-local resistance remain unaffected due to chiral transport. To calculate quantum localization correction, we need to calculate  the resistances from probabilities neglecting the phase acquired via multiple scattering off disordered contacts. The conductance matrix $G$ built from transmission and reflection probabilities is-
{
\begin{equation} 
 G=\frac{2e^2}{h}\frac{1}{b''}\left(\begin{smallmatrix}
(1-R^5)T&-T^2R^4&-T^2R^3&-T^2R^2&-T^2R&-T^2\\
 -T^2&(1-R^5)T&-T^2R^4&-T^2R^3&-T^2R^2&-T^2R\\
 -T^2R&-T^2&(1-R^5)T&-T^2R^4&-T^2R^3&-T^2R^2\\
 -T^2R^2&-T^2R&-T^2&(1-R^5)T&-T^2R^4&-T^2R^3\\
 -T^2R^3&-T^2R^2&-T^2R&-T^2&(1-R^5)T&-T^2R^4\\
 -T^2R^4&-T^2R^3&-T^2R^2&-T^2R&-T^2&(1-R^5)T\end{smallmatrix}\right),
 \end{equation}
}where $b''=(1-R^6)$. Again, setting the current through voltage probes $2,3,5,6$ equal to zero, and choosing the reference potential $V_4=0$ we get the potentials $V_2$ and $V_4$ in terms of $V_1$. We get the Hall resistance $R^{}_H$, 2-terminal resistance $R^{}_{2T}$, and non-local resistance $R^{}_{NL}$ via probabilities as-
\begin{eqnarray}
R^{}_H&=&\frac{h}{2e^2},
\text{ } R^{}_{2T}=\frac{h}{2e^2}\frac{(1+D)}{(1-D)},\text{ } R_L^{}=R^{}_{NL}=0.
\end{eqnarray}
Increasing the number of terminals from four to six we see no change attributed to the derived resistances via probabilities. The quantum localization correction for Hall, longitudinal, 2-terminal and non-local resistances in the six terminal QH sample are thus with $R_X^{QL}=\langle R_X^{Amp}\rangle-R_X^{}$, with $X=H, L, 2T, NL$-
\begin{eqnarray}
R^{QL}_H&=&\frac{h}{2e^2}\frac{2D^6}{1-D^6},\quad R^{QL}_{2T}=\frac{h}{2e^2}\frac{2D^6(1+D)}{(1-D)(1-D^6)},\nonumber\\
R^{QL}_{L}&=&0,\qquad R^{QL}_{NL}=0.
\end{eqnarray}
\begin{figure}
\includegraphics[width=0.47\textwidth]{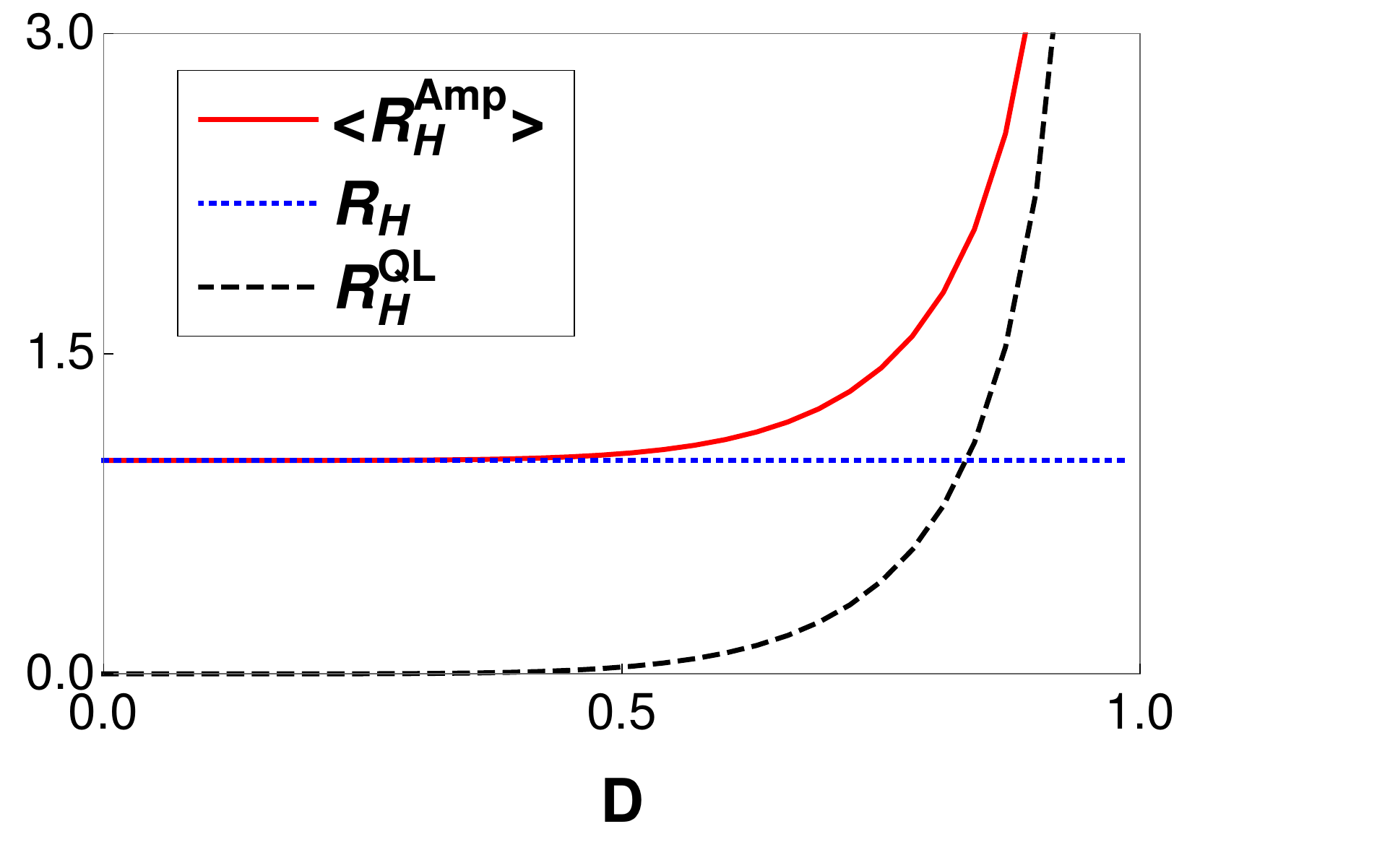}
\includegraphics[width=0.45\textwidth]{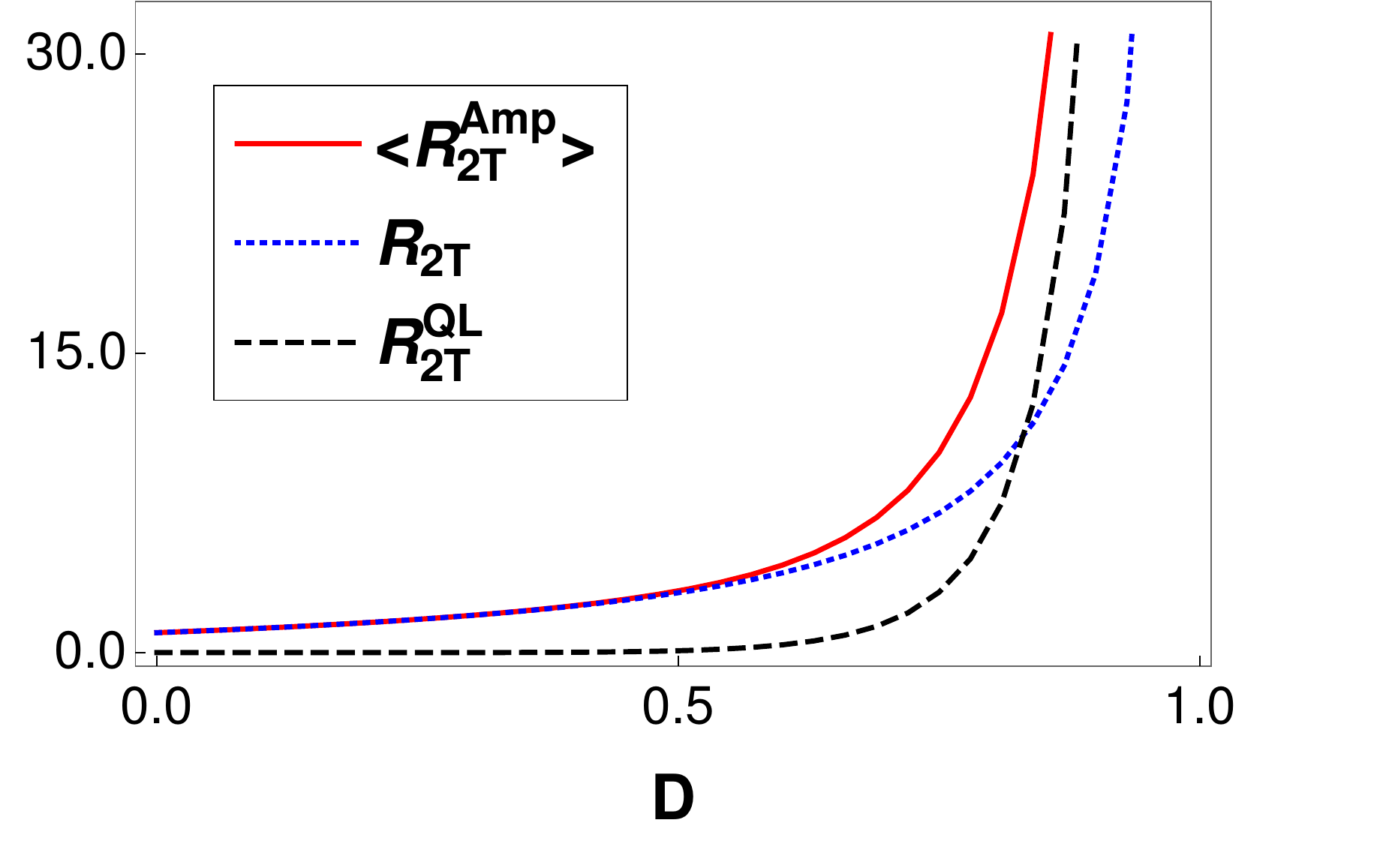}
\vspace{0cm}\\
\text{\qquad\qquad\qquad(a)\qquad\qquad\qquad\qquad\qquad\qquad\qquad\qquad(b)\qquad\qquad\qquad}
\caption{(a) Hall resistance and (b) 2-terminal resistance in units of $h/2e^2$ via scattering amplitudes, via probabilities and the quantum localization correction plotted versus $D$ the  disorderedness of a contact, assuming each contact to be equally disordered.}
\end{figure}
Again for equally disordered contacts the $R_L^{QL}=0$. One can check for the case wherein all contacts at upper edge are disordered with strength $D_u=D_1=D_2=D_3$ while lower contacts are disordered with strength $D_l=D_4=D_5=D_6$ then also $\langle R^{Amp}_L\rangle=R_L=0$. Further when all contacts are unequally disordered, i.e., $D_1\ne D_2\ne D_3\ne D_4\ne D_5\ne D_6$ then also $\langle R^{Amp}_L\rangle=R_L=0$. Thus, there is no localization correction for longitudinal and non-local resistances for QH edge
modes. This quantum localization correction depends on the strength of disorder unlike the
weak localization correction seen in quantum diffusive transport. The quantum localization
correction for partially disordered sample, say $D\leq(1/2)$ is very small and negligible, only
for $D>(1/2)$ does the correction become significant. In Fig.~4 we show the quantum localization correction (along-with the resistance calculated from Probabilities and the phase averaged resistance calculated via the amplitudes) to the Hall resistance and the 2-terminal resistance in a six terminal quantum Hall system as function of the contact disorder strength $D$, assuming each contact to be equally disordered. 
\begin{figure*}
  \centering \subfigure[]{ \includegraphics[width=0.33\textwidth]{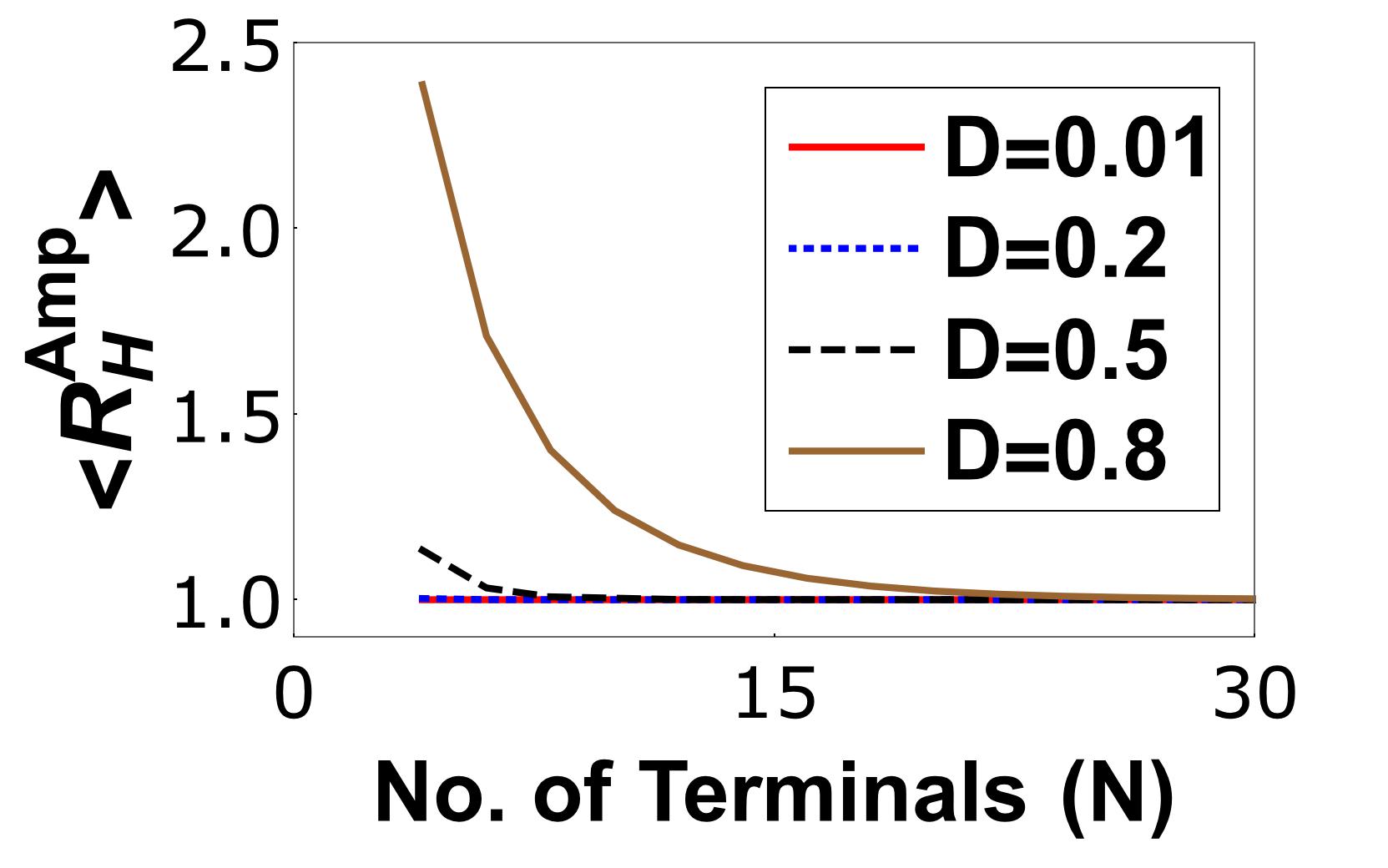}}
 \centering    \subfigure[]{ \includegraphics[width=0.32\textwidth]{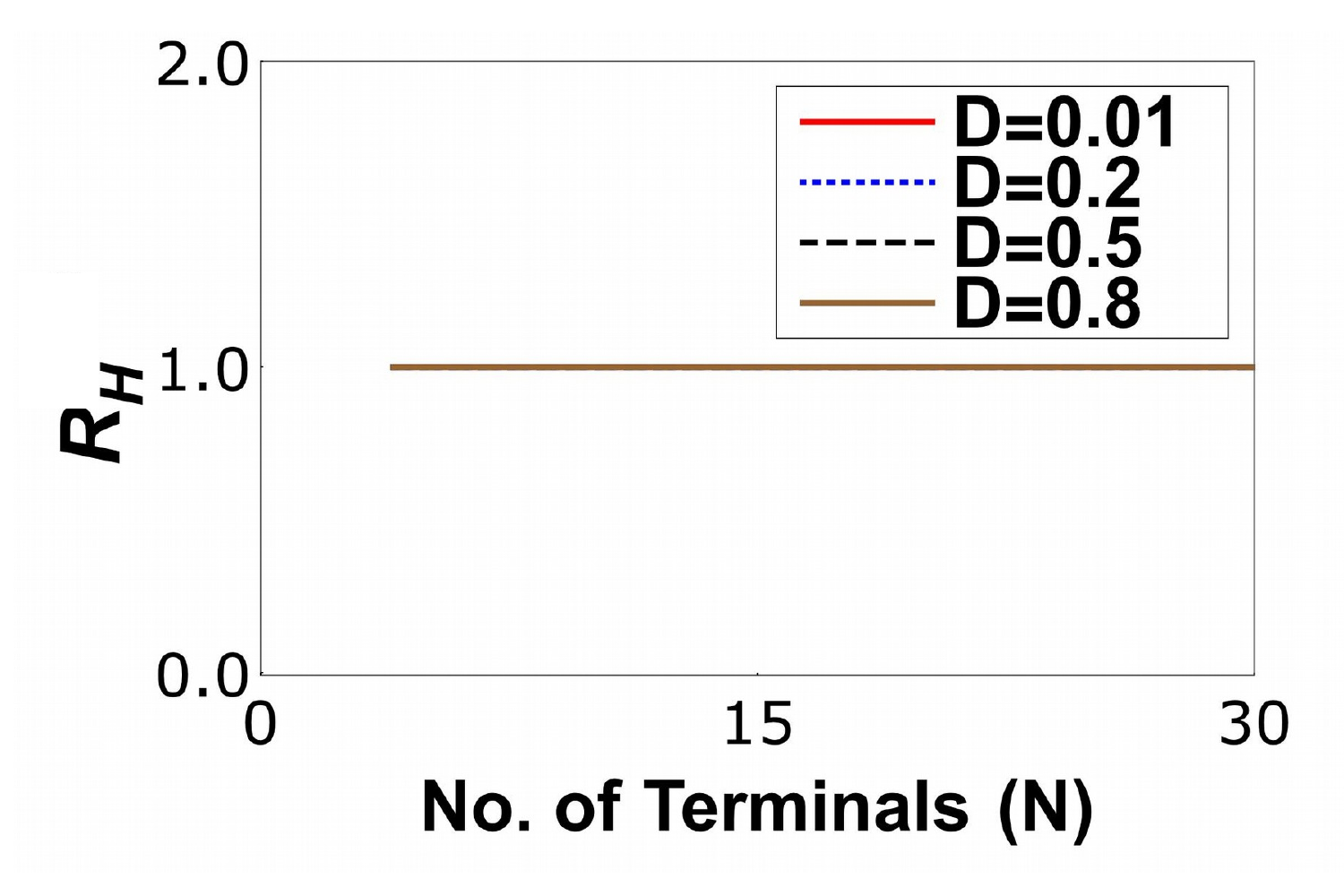}}
 \centering \subfigure[]{\includegraphics[width=.33 \textwidth]{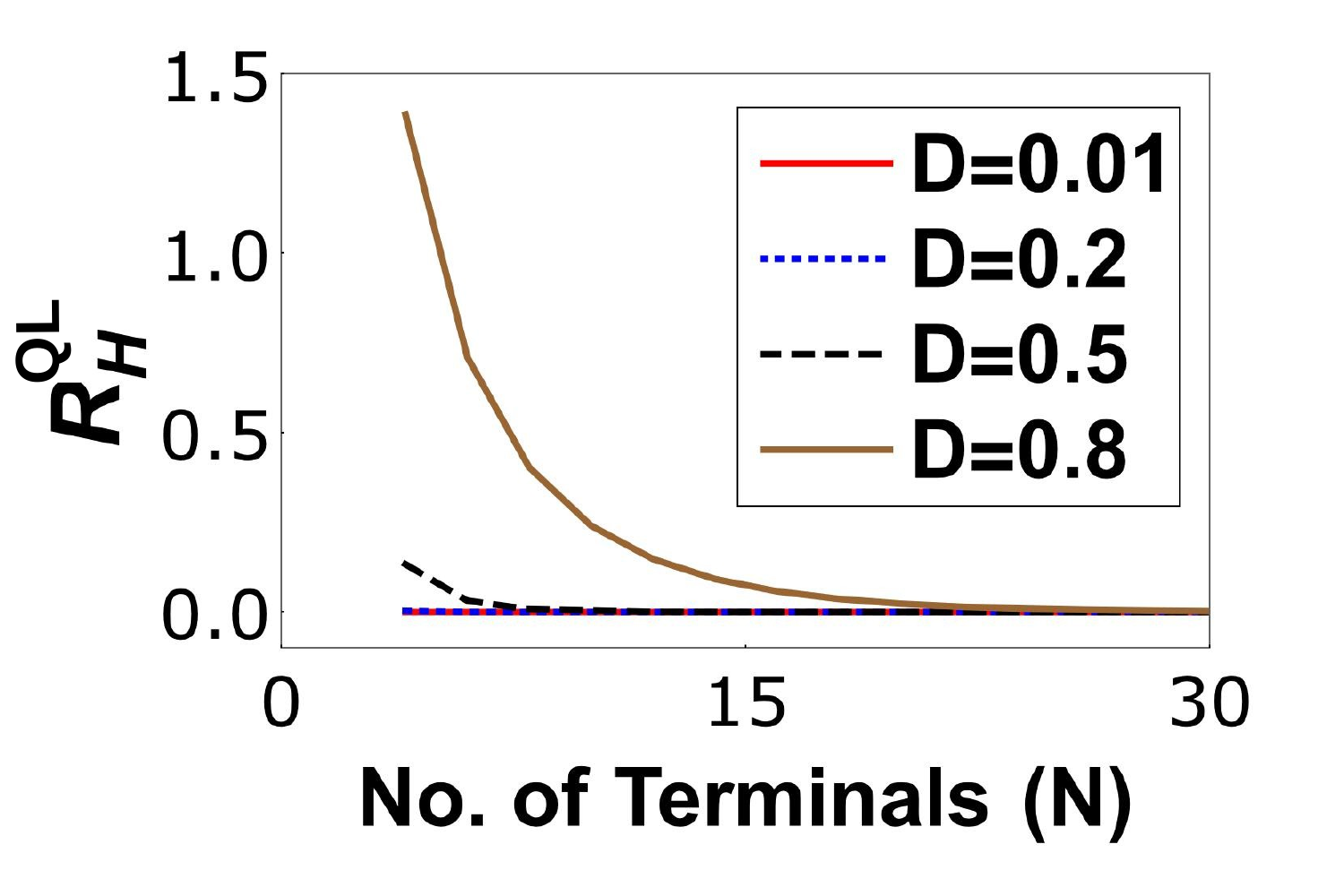}}
\vskip -0.2 in \caption{Hall resistance in units of $\frac{h}{2e^2}$ in a N terminal Quantum Hall bar with all contacts equally disordered for (a) via scattering amplitude, (b) via probabilities, and (c) quantum localization correction.}
\end{figure*}

\begin{figure*}
  \centering \subfigure[]{ \includegraphics[width=0.33\textwidth]{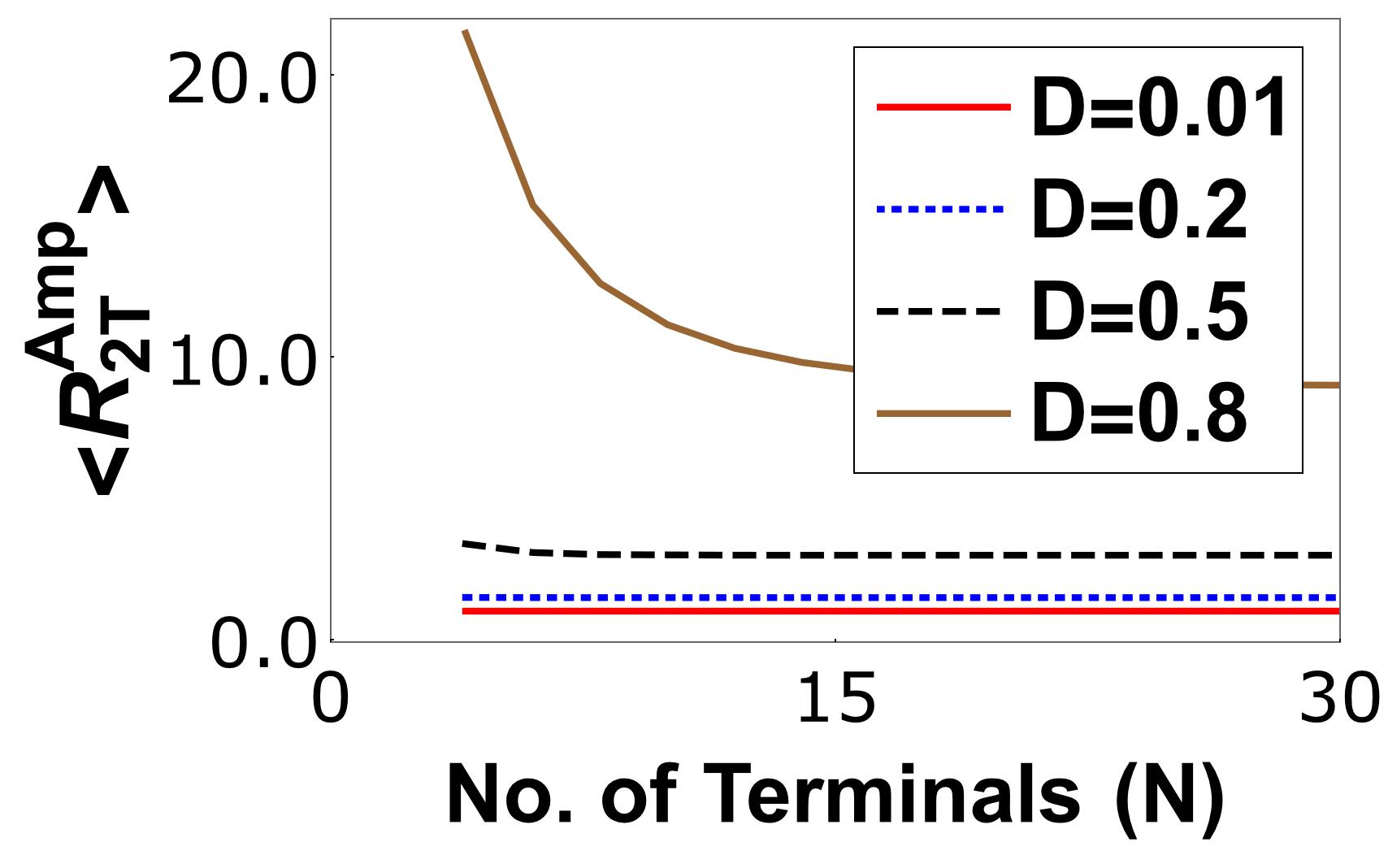}}
 \centering    \subfigure[]{ \includegraphics[width=0.32\textwidth]{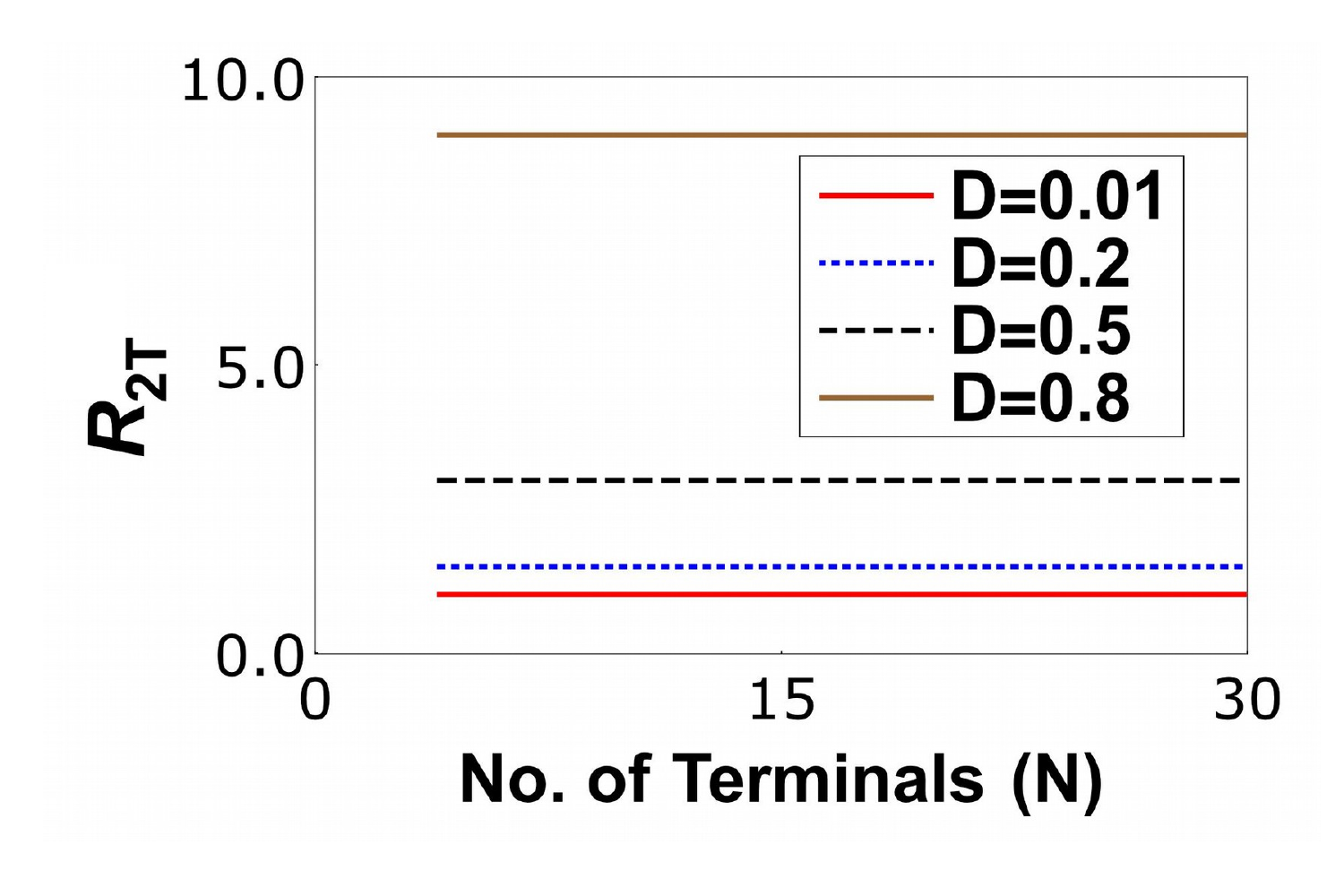}}
 \centering \subfigure[]{\includegraphics[width=.33 \textwidth]{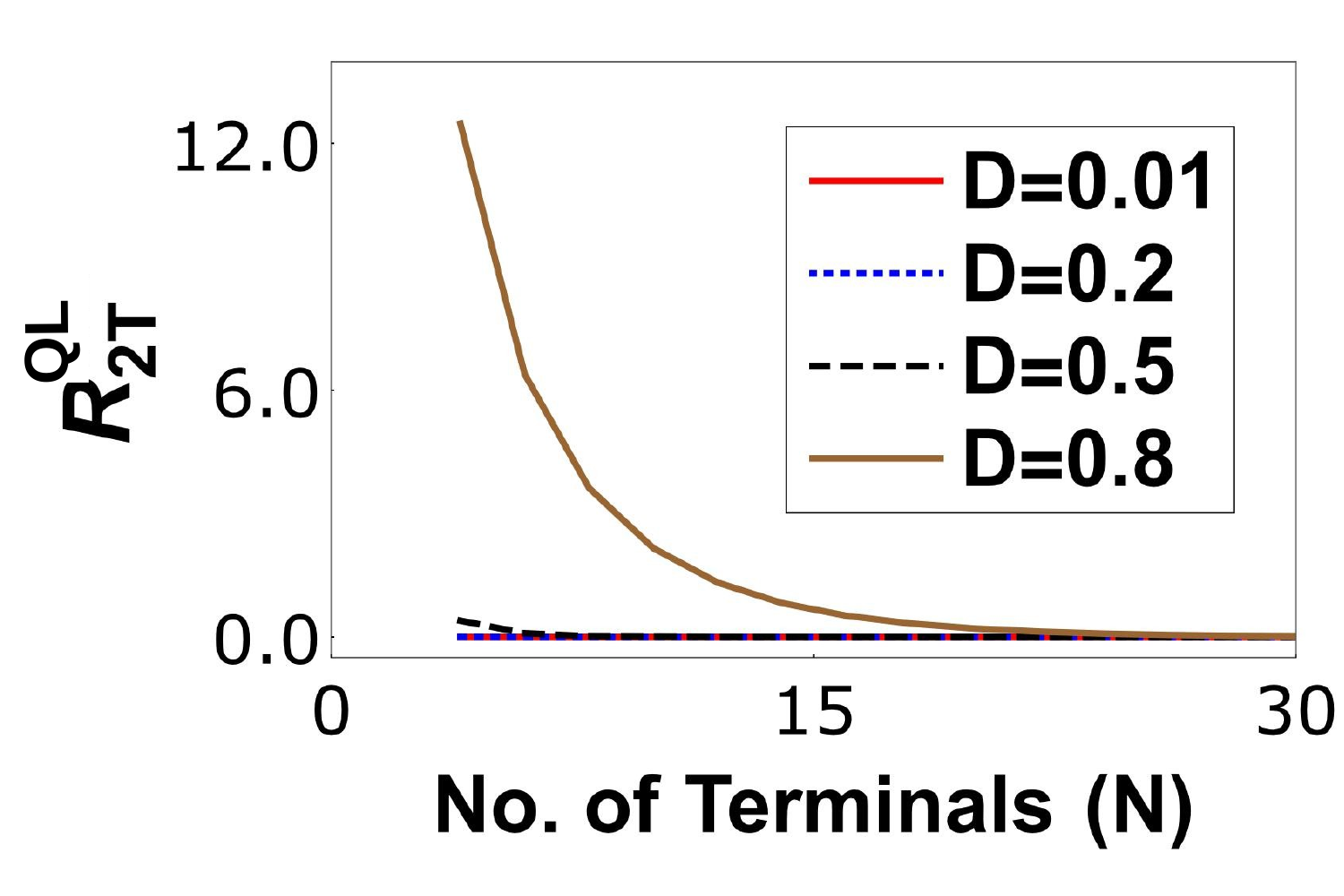}}
\vskip -0.2 in \caption{2-terminal resistance in units of $\frac{h}{2e^2}$ in a N terminal Quantum Hall bar with all contacts equally disordered for (a) via scattering amplitude, (b) via probabilities, and (c) quantum localization correction.}
\end{figure*} 

 \subsection{N terminal system with all contacts disordered}
An N terminal QH system is shown in Fig.~3(c) with all contacts equally disordered. Contacts $1$ and $k$ are current probes and contacts $2, 3,...k-1,k+1,...N$ are voltage probes, so that, current through these contacts are set to zero. Thus $I_2=I_3=....=I_{k-1}=I_{k+1}=....=I_N=0$. The scattering matrix for the system in Fig.~3(c) is shown below-
 {
 \begin{equation} 
 S=\frac{1}{c}\left(\begin{smallmatrix}
 (r-r^{N-1}e^{\phi})e^{\phi_1}&...&-t^2r^{N-k}e^{i\phi_{k(k+1)..1}}&...&-t^2e^{i\phi_{1N}}\\
 .&...&.&...&.\\
  .&...&.&...&.\\
   -t^2r^{k-2}e^{i\phi_{12..k}}&...& (r-r^{N-1}e^{\phi})e^{\phi_k}&...&-t^2r^{k-1}e^{i\phi_{N12..k}}\\
    .&...&.&...&.\\
     .&...&.&...&.\\
   -t^2r^{N-2}e^{i\phi_{12..N}}&...&-t^2r^{N-k-1}e^{i\phi_{k(k+1)..N}}&...&(r-r^{N-1}e^{\phi})e^{i\phi_{N}}
\end{smallmatrix}\right),
 \end{equation}
}where $c=1-r^Ne^{i\phi}$, $r=r_i$ and $t=t_i$ are the reflection and transmission amplitudes at contact $i$ for $i=1,..,N$. Here, in Eq.~(18) to calculate the scattering matrix we have chosen all equally disordered contacts since for unequally disordered contacts it is easy to calculate the scattering matrix but difficult to write in compact fashion. In Eq.~(18), $\phi=\phi_1+\phi_2+..+\phi_N$, is the sum of all scattering phases acquired at each disordered contact and $\phi_{ij..k}=\phi_i+\phi_j+..+\phi_k$. Unitarity of the scattering matrix $S$ implies the conservation of current in the system. The conductance matrix $G$ of the system derived from the scattering matrix $S$ in Eq.~(18), following Eq.~(1), is-
{
\begin{equation} 
 G=\frac{1}{c'}\left(\begin{smallmatrix}
 T(1-R^{N-1})&...&-T^2R^{N-k}&...&-T^2\\
   .&...&.&...&.\\
     .&...&.&...&.\\
   -T^2R^{k-2}&...&T(1-R^{N-1})&...&-T^2R^{k-1}\\
   .&...&.&...&.\\
     .&...&.&...&.\\
   -t^2R^{N-2}&...&-T^2R^{N-k-1}&...&T(1-R^{N-1})
\end{smallmatrix}\right),
 \end{equation}
}where $c'=1+R^{N}-2R^{N/2}\cos\phi$, $T=1-D=|t|^2$ and $R=D=|r|^2$. Since the current through voltage probes $2,3,...,k-1,k+1,...,N$ are zero, so $I_2=I_3=...=I_{k-1}=I_{k+1}=...=I_N=0$, and choosing reference potential $V_k=0$ we get the potentials $V_2$, $V_3$,..., $V_{k-1}$, $V_{k+1}$,...and $V_N$ in terms of $V_1$. Thus, the Hall resistance $R^{Amp}_H=R_{1k,2N}=\frac{(V_2-V_N)}{I_1}$, 2-terminal resistance $R^{Amp}_{2T}=R_{1k,1k}=\frac{(V_1-V_k)}{I_1}$, longitudinal resistance $R^{Amp}_L=R_{1k,23}=\frac{(V_2-V_3)}{I_1}$ and non-local resistance $R^{Amp}_{NL}=R_{12,(k+1)k}=\frac{(V_{k+1}-V_k)}{I_1}$ (to calculate the non-local resistance only we have to consider contacts $1,2$ are current probes and contacts $3,4,..,k-1,k,k+1,...,N$ are voltage probes) becomes-
\begin{eqnarray}
R^{Amp}_H&=&\frac{h}{2e^2}\frac{1+D^N-2D^{N/2}\cos\phi}{1-D^N},\nonumber\\
R^{Amp}_{2T}&=&\frac{h}{2e^2}\frac{(1+D)(1+D^N-2D^{N/2}\cos\phi)}{(1-D)(1-D^N)},\nonumber\\
R^{Amp}_L&=&R^{Amp}_{NL}=0.
\end{eqnarray}
After averaging over the phase shift $\phi$ acquired by the electron via scatterings at the disordered contacts, we get-
 \begin{eqnarray}
\langle R^{Amp}_H\rangle&=&\frac{h}{2e^2}\frac{1+D^N}{1-D^N},\nonumber\\
\langle R^{Amp}_{2T}\rangle&=&\frac{h}{2e^2}\frac{(1+D)(1+D^N)}{(1-D)(1-D^N)},\nonumber\\
\langle R^{Amp}_L\rangle&=&\langle R^{Amp}_{NL}\rangle=0.
\end{eqnarray}
To calculate quantum localization correction, we need again to calculate the resistances using the probabilities ignoring the phase acquired by the electron. The conductance matrix $G$ then is-
  \begin{equation} 
 G=\frac{1}{c''}\left(\begin{smallmatrix}
 T(1-R^{N-1})&...&-T^2R^{N-k}&...&-T^2\\
   .&...&.&...&.\\
     .&...&.&...&.\\
   -T^2R^{k-2}&...&T(1-R^{N-1})&...&-T^2R^{k-1}\\
   .&...&.&...&.\\
     .&...&.&...&.\\
   -t^2R^{N-2}&...&-T^2R^{N-k-1}&...&T(1-R^{N-1})
\end{smallmatrix}\right),
 \end{equation}
 where $c''=(1-R^{N})$. Again, setting the current through voltage probes $2,3,...,k-1,k+1,...,N$ equal to zero, and choosing the reference potential $V_k=0$ we get the potentials $V_2$, $V_3$, $V_{k-1}$, $V_{k+1}$ and $V_N$ in terms of $V_1$. We get the Hall resistance $R^{}_H$, 2-terminal resistance $R^{}_{2T}$, and non-local resistance $R^{}_{NL}$ via probabilities as-
 \begin{eqnarray}
R^{}_H&=&\frac{h}{2e^2},
\text{ } R^{}_{2T}=\frac{h}{2e^2}\frac{(1+D)}{(1-D)},\text{ } R_L^{}=R^{}_{NL}=0.
\end{eqnarray}
The quantum localization corrections for Hall, longitudinal, 2-terminal and non-local resistances in the N terminal QH sample are-
\begin{eqnarray}
R^{QL}_H&=&\frac{h}{2e^2}\frac{2D^N}{1-D^N},\quad R^{QL}_{2T}=\frac{h}{2e^2}\frac{2D^N(1+D)}{(1-D)(1-D^N)},\nonumber\\
R^{QL}_{L}&=&0,\qquad R^{QL}_{NL}=0.
\end{eqnarray}
\begin{figure*}
 \centering {\includegraphics[width=0.45\textwidth]{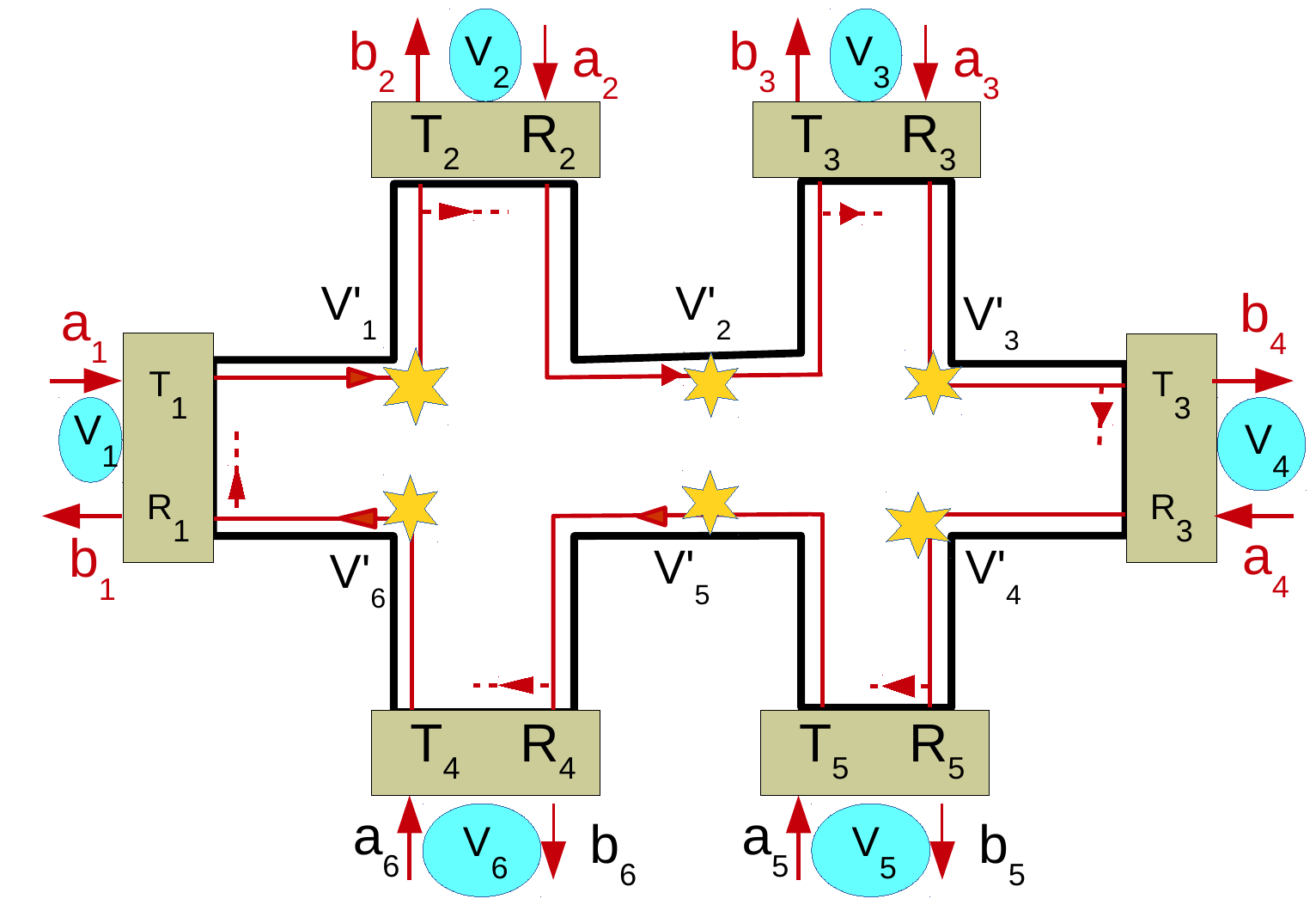}}
 \vspace{-.3cm}
\caption{Six terminal QH bar with all contacts disordered and inelastic scattering.}
\end{figure*}
This completes our analysis of QH system with disorder. The plots for the resistances and quantum localization correction are shown in Figs.~5, 6 for Hall and 2-terminal resistance. There is no quantum localization correction for longitudinal or non-local resistance. So, we do not plot these. We see from Fig.~5(a), that Hall resistance derived from scattering amplitudes decreases  with number of terminals while the Hall resistance derived from probabilities is quantized and unaffected by the number of terminals, see Fig.~5(b). In Fig.~5(c), we see that the quantum localization correction for Hall resistance decreases with increasing number of contacts. We also see that large amount of disorder at the contacts leads to large quantum localization correction. In Figs.~6(a,b,c), we see that the 2-terminal quantum localization correction too decreases with increasing number of contacts. For large number of contacts ( $N > 15$ ) regardless of disorder the resistances calculated considering phase and neglecting phase are identical. Only for small number of terminals and disorder strength $D > 1/2$ is the difference between resistances (2-terminal and Hall) when considering phase and ignoring phase is substantial.
\section{Effect of inelastic scattering on quantum localization correction}
In previous sections, we discussed the quantum localization correction to the QH edge modes in presence of all disordered contacts. Now we introduce inelastic scattering within the system along with the disordered contacts to see its effect on the quantum localization correction. In quantum-diffusive transport regime it has been shown that inelastic scattering completely kills the weak localization correction\cite{datta}. We want to see the effect of inelastic scattering on  the quantum localization correction addressed in previous sections. A six terminal QH bar with all disordered contacts including inelastic scattering is shown in Fig.~7. When the length between the disordered contacts is larger than the phase coherence length of the electronic edge modes, then electrons coming from two different contacts equilibrate their energy and their population in the edge modes via inelastic scattering (shown by the yellow starry blobs in Fig.~7). In Fig.~7, starry blobs are shown at particular places between two contacts, however, it does not mean that inelastic scattering is happening only at those places, it can can happen anywhere between the two contacts. \\
Once the edge modes are equilibrated via inelastic scattering to a common potential $V_i'$ ($i=1-6$), they remain equilibrated throughout the length between two contacts. Edge modes coming from two different contacts have different potentials and they are equilibrated to a common potential via inelastic scattering mediated by electron-electron interaction or electron-phonon interaction. In presence of inelastic scattering there are no longer multiple paths from one contact to another. This can be understood in this way- lets say an electron coming out of contact $1$ can arrive at equilibrating potential $V_1'$ by following only one path and then it loses its phase via equilibration of energy with other electrons. So there is no way it can reflect back to contact $1$ again with the same energy. An electron suffers both elastic and inelastic scattering. Elastic scattering occurs at the contacts via impurities and generates a phase shift to the electrons initial phase. While inelastic scattering occurs between contacts leading to not only loss of phase acquired but also a change in energy of the electron from $V_1$ to $V_1'$. This loss of phase and change in energy of electrons occurs inelastically at each of the starry blobs and finally the electron comes back with an energy $V_6'$ and without any phase memory. Thus the multiple path at fixed energy seen with only elastic scattering at the contacts is no longer possible. First because electron loses phase memory and second its energy regularly changes due to inelastic scattering. As there are no multiple paths from one contact to another, the transmission probability derived from scattering amplitudes and that derived from probabilities are same. So, there will be no quantum localization correction in presence of inelastic scattering. Using probabilities the resistances are already derived in our previous works, see Refs.~\cite{arjun,arjun1}, as- the Hall resistance $R_H=\frac{h}{2e^2}$, longitudinal resistance $R_L=0$, 2-terminal resistance $R_{2T}=\frac{h}{2e^2}\frac{(1-D_1D_4)}{(1-D_1)(1-D_4)}$ and non-local resistance $R_{NL}=0$.

\section{Conclusion}
We see that resistances in quantum Hall systems are affected by a quantum localization correction if and only when all contacts in a QH sample are disordered. The local or 2-terminal and Hall resistance are seen to be affected by the quantum localization correction while longitudinal and non-local resistance remain unaffected by it due to chiral transport. On the other hand, in presence of inelastic scattering this quantum localization correction vanishes.

\acknowledgments
This work was supported by funds from SCIENCE \& ENGINEERING RESEARCH BOARD, DST, Government of India, Grant No.   EMR/2015/001836.

\end{document}